\documentclass[10pt]{article}
\usepackage{amsmath, amsthm}
\usepackage{amsmath}
\usepackage{amsfonts}
\usepackage{amssymb}
\usepackage{verbatim}
\usepackage{amscd}
\usepackage{amsthm}

\newtheoremstyle{thm}{1.5ex}{1.5ex}{\itshape\rmfamily}{}
{\bfseries\rmfamily}{}{2ex}{}

\newtheoremstyle{rem}{1.3ex}{1.3ex}{\rmfamily}{} 
{\itshape}
{} {1.5ex}{}

\newtheoremstyle{theorem}{1.5ex}{1.5ex}{\itshape\rmfamily}{} {\bfseries\rmfamily}{}{2ex}{}

\newtheoremstyle{def}{1.5ex}{1.5ex}{\slshape\rmfamily}{} {\bfseries\rmfamily}{}{2ex}{}

\newtheoremstyle{rem}{1.3ex}{1.3ex}{\rmfamily}{} {\itshape}
{} {1.5ex}{}

\theoremstyle{theorem}
\newtheorem{theorem}{Theorem}[section]

\newtheorem{proposition}[theorem]{Proposition}

\newtheorem*{Main Theorem}{Main Theorem.}
\newtheorem{corollary}[theorem]{Corollary}
\newtheorem*{special theorem}{Lindeberg-Feller Theorem for Martingales}

\theoremstyle{def}

\theoremstyle{rem}
\newtheorem{remark}{{\itshape Remark}}[]

\numberwithin{equation}{section}

\begin{document}

\title
{\Large{Percolation Phenomena in Low and High Density Systems}}

\author
{\large{L. Chayes$^1$}  J.L. Lebowitz$^{2,3}$ and V. Marinov$^3$}
\maketitle

\vspace{-4mm}
\centerline{${}^1$\textit{Department of Mathematics, UCLA, Los Angeles,
CA}}
\centerline{${}^2$\textit{Department of Mathematics, Rutgers University, New Brunswick, NJ}}
\centerline{${}^3$\textit{Department of Physics, Rutgers University, New Brunswick, NJ}}

\begin{quote}
{
\footnotesize {\bf Abstract:}
We consider the 2D quenched--disordered $q$--state Potts ferromagnets and show that at self--dual points any amalgamation  of $q-1$ species will fail to percolate despite an overall (high) density of $1-q^{-1}$.  Further, in the dilute bond version of these systems, if the system is just above threshold, then throughout the low temperature phase there is percolation of a single species despite a correspondingly small density.  Finally, we demonstrate both phenomena in a single model by considering a ``perturbation'' of the dilute model that has a self--dual point.  We also demonstrate that these phenomena occur, by a similar mechanism, in a simple coloring model invented by O. H\"aggstr\"om.
} 
\end{quote}

\section{\large{Introduction}}

The purpose of this note is to address the question of whether there are ``natural'' translation and (lattice) rotation invariant ergodic measures on configurations $\eta \in
 {\{0,1\}}^{\mathbb Z^d}$, $d \geq 2$, for which site percolation occurs when the density of occupied sites, $\rho = \langle \eta_i \rangle$, is very close to zero and/or fails to occur if the density is close to one. It is known that neither phenomenon occurs for systems with product measures (independent percolation) \cite{GHM,GrimBook}, where, in most situations, there is a sharp percolation transition when $\rho^{-1}$ is of the order of the lattice coordination number (e.g.~$2d$).   
 The question then is what happens when the occupancy of different sites is specified in some natural way which is {\it not} independent. 
Of particular interest are equilibrium spin--systems and non-equilibrium stationary states of interacting particle systems such as  
the voter model, the contact process and their various generalizations~\cite{Lig}. 
Furthermore, one may also consider projections of such measures, i.e.~given a measure on $[\Omega]^{\mathbb Z^d}$, where $\sigma_i \in \Omega$ is the set of possible states at $i \in  \mathbb Z^d$, we define an occupation variable $\eta_i = 0,1$ , with $\eta_i=1$ if 
$\sigma_i \in \Omega_1$ and $\eta_i = 0$ if $\sigma_i  \in \Omega \setminus \Omega_1$. Well known examples are the fuzzy q-state Potts model with $q=r+s$ and $\eta_i=1$ if $\sigma_i \in (1,..,r)$, and systems where $\sigma_i \in \mathbb R$, and $\eta_i =1 $ if $\sigma_i \geq h$ and zero otherwise ~\cite{Chayes,Hgg3,LM}. 

Adams and Lyons \cite{AL} considered these types of question for measures on homogeneous trees that are invariant under the graph automorphisms.  They found that whenever the density is large enough there indeed is percolation, see also~\cite{Hgg1} and~\cite{BLPS}. 
Such a result does {\it not} hold for ${\mathbb Z^d}$; it is easy to construct counterexamples which exploit, in an obvious way, the vanishing surface to volume ratio of regular sets \cite{GHM}.
Also, for $d=2$ one may consider the infinite cluster of supercritical independent percolation near threshold and simply declare the complimentary sites to have $\eta_1 \equiv 1$; these complimentary sites will have density close to one and fail to percolate.  Higher dimensional analogs may even be possible, albeit at different thresholds (c.f.~\cite{ACCFR}) and the converse phenomenon can also be constructed by these means:  Consider a slightly super--critical percolating cluster and declare its compliment to be vacant ($\eta_i \equiv 0$).  While such examples are not particularly interesting in their own right, they demonstrate the existence of mechanisms for these phenomena which might occur in more realistic systems.  

Along these lines, a more interesting set of examples concern the so--called divide and color (DaC) models invented by H\"aggstr\"om \cite{Hgg2}.  As we will show, for these models in $d = 2$ there are no density limitations for the absence or presence of percolation. But even the DaC models are a bit artificial and one should ask whether there are more natural examples where such phenomena can occur.

In this note, the principal system under study is the 2D disordered Potts ferromagnet.  First we will demonstrate that at self--dual critical points, there is an absence of percolation despite a density which may be arbitrarily close to one.  The complimentary result, namely percolation at small density would follow immediately from continuity of the magnetization -- a currently open and challenging problem.  (We will state some {\it inconclusive} results concerning this question.)  However, almost trivially, the diluted versions of the disordered Potts models exhibit this property throughout the low temperature phase (i.e.~when the magnetization is positive) provided the media itself is near threshold; we will provide a formal proof.  Finally we will combine the two sets of results via the consideration of models with {\it strong} and {\it weak} interactions.  These models inherit enough of the features of the dilute model to display the small density percolation property yet, unlike the dilute models, can have a self--dual point.  

Most of the above stated results are proved in the context of a fixed realization of the disorder.  However, since any particular realization is not translation invariant we do not satisfy the naturalness criterion that was stated at the beginning of this section.  Therefore we will consider the entire problem from the perspective of the {\it quenched} measures.  These objects are manifestly translation invariant and we will demonstrate various other amenable properties, such as strong mixing. 

The remainder of this paper is organized as follows:  In section 2 we define the DaC models and, thereafter, we state and prove our results for this system.
In section 3, we define the general quenched disordered Potts ferromagnets but with the emphasis on $d = 2$.  We define the {\it quenched measures} for these systems; certain general properties of these quenched measures are stated (with the proofs postponed).  We then prove our main results:  (i)  At certain self--dual critical points, any one of the species is sufficient to prevent percolation of the combined efforts of all the others. (ii)  For the dilute model, with the active bonds in slight excess of the percolation threshold, there is a non--trivial low temperature phase characterized by the percolation of just one of the species notwithstanding (for $q \gg 1$) the paucity of the overall density of this species.  Finally we construct a weak--bond/strong--bond disordered model with a self--dual point {\it and}, at least in the vicinity of the transition temperature, a regime of percolation at low density.  
In the next section, we will provide some discussion and finally, in the appendix, we prove the stated properties of the quenched measures in the disordered Potts systems.    
For further background discussion, see~\cite{GHM} Section 5.2.
\section{\large{Critical 2D DaC Models}}
\subsection{\large{Definitions}}  On  the ``graph'' $\mathbb Z^2$ let 
$p \in [0,1]$, $q$ an integer $\geq 2$ and let $\lambda_1, \dots \lambda_q$ be in $(0,1)$ with $\sum_q \lambda_q =1$.  The relevant measure, which we denote by $\mu_{p,\underline \lambda}$ (with $\underline \lambda = (\lambda_1, \dots \lambda_q)$)
is defined as follows:  Each edge is {\it occupied} (or {\it vacant}) independently with probability $p$ (and $(1-p)$).  For each given configuration of edges, the finite connected clusters of sites are independently ``colored'' in one of $q$ ways with respective probabilities $\lambda_1, \dots \lambda_q$.  If the bonds do {\it not} percolate, then the model has been defined.  If there is percolation, $p > p_{c} = \frac12$, one can define $q$ distinct ``extremal'' measures, $\mu^{[j]}_p(\lambda_1...\lambda_q)$, where the infinite cluster is assigned the jth color and convex combinations may be chosen according to any desired prescription.  The extreme measures can be justified/constructed via the limit of finite volume arrangements wherein the boundary of the region is deemed to be of the particular color chosen.  

Despite its somewhat artificial appearance, the DaC model has some interesting features (not discussed in the present work) and in fact can actually be realized in a number of circumstances.  In particular, for $\lambda_{1} = \dots = \lambda_{q}$ it is the zero temperature limit of the random diluted $q$--state Potts model that will be featured in the next section.  Further, if one constructs the $q$--state diluted {\it voter model}, with an initial distribution that is {\it a priori} equal and independent it is not hard to see (e.g.~by the methods of \cite{Lig} Chapter V) that the infinite time limit of this setup is precisely the DaC with corresponding $p$ and $\lambda_{1} = \dots = \lambda_{q}$.


\subsection{\large{Statements and Proofs}}
Our first theorem, the key parts of which are actually contained in \cite{Hgg2}, reads as follows:

\begin{theorem}
\label{DaC1}
For all $\lambda_{1} \in (0,1)$, the measures 
$\mu_{p_{c},(\lambda_1,\lambda_2)}$ 
have, with probability one, no (connected) site percolation or $\ast$--\thinspace connected site percolation for either of the species.
\footnote
{$\ast$--percolation means an infinite cluster under the relaxed rules that neighbors or next nearest neighbors are considered connected.} 
\end{theorem}
\begin{remark}
We remark that the case of interest in the present work is when one of the $\lambda$'s is small; the setup in \cite{Hgg2} was the completely symmetric $p = \lambda_1 = \lambda_2 = \frac{1}{2}$.  We further remark that by the obvious method of aggregating colors, the above extends to versions of the DaC with more colors.  In particular, with $q$ species, the union of any of the $q-1$ species is denied percolation by the single remaining color.
\end{remark}
\begin{proof}
The critical bond percolation model on $\mathbb Z^2$ has the key feature that there are infinitely many closed circuits of occupied bonds surrounding the origin 
{\it and} infinitely many closed circuits of dual bonds -- edges on $(\mathbb Z + \frac{1}{2})^{2}$ -- traversal to the vacant bonds surrounding the origin ~\cite{GrimBook}.  
Since the latter separate the former, it is evident that there are infinitely many 
{\it disjoint} and {\it separated} occupied bond circuits surrounding the origin each of which belong to clusters of color  type 1 or color type 2 with respective probabilities 
$\lambda_{1}$ and $\lambda_{2}$. This rules out the possibility of the origin belonging to an infinite connected or even $\ast$--connected cluster of either type.
\end{proof}
The above theorem is complimented by the situation above threshold:
\begin{theorem}
\label{above}
For every $\epsilon >0$ there is a region of $p>p_c$ and $\lambda_1$ such that $\mu^{[1]}_{p,(\lambda_1, \lambda_2)}$ has percolation of the first color but the overall density of the first color is less than $\epsilon$.
\end{theorem}
\begin{proof}
Consider the measure $\mu^{[1]}_{p,(\lambda_1, \lambda_2)}$ with $p$ (only slightly) in excess of $p_c$.   
It is clear, by conditioning on the events that a particular site does and does not belong to the infinite bond cluster, that the color 1 density is given by $P_\infty(p) + \lambda_1(1 - P_\infty(p))$.  Here $P_\infty(p)$ denotes the fraction of {\it sites} belonging to the infinite cluster of the bonds.  
Since, for the 2d bond problem, $P_\infty(p)$ is known to vanish continuously \cite{Russo} the result follows.
\end{proof}
\section{Disordered Potts Ferromagnets}
\subsection{\large{Background Considerations}}
We let $\{J_{i,j}\mid i,j \text{ neighbors in } \mathbb Z^2\}$
denote a collection of iid non--negative random variables with common distribution $F$ not concentrated at a single point; we emphasize that we may allow $J_{i,j} = 0$ with non--zero probability, a case we will refer to as {\it dilution}.  
The bond--random Potts model is defined, for fixed realization of the $(J_{i,j})$, by the formal Hamiltonian
\begin{equation}
\label{FH}
-H  =  -H_{\mathbb J}  =  \sum_{\langle i,j\rangle}J_{i,j}\delta_{\sigma_i,\sigma_j}
\end{equation}
where $\langle i,j\rangle$ denotes nearest neighbors, $\mathbb J$ represents the collection $(J_{i,j})$ and $\sigma_i \in \{1, \dots q\}$ are the spin--variables associated with the sites of $\mathbb Z^2$.  Most often, we will absorb the usual temperature parameter into the definition of $\mathbb J$.

For fixed $\mathbb J$ in a finite volume $\Lambda$, the Gibbs measures with specified boundary conditions are defined in the usual fashion.  E.g.~if $\Lambda \subset \mathbb Z^2$ and the boundary $\partial \Lambda$ has fixed boundary spins 
$\sigma_{\partial \Lambda}$, we may write
\begin{equation}
\label{Gibbs weights}
\mathbb P^{\sigma_{\partial \Lambda}}_{\mathbb J; \Lambda}
(\sigma_{\Lambda})
 = 
\frac{e^{-H_{\mathbb J}(\sigma_{\Lambda}\mid \sigma_{\partial \Lambda})}}{Z^{\sigma_{\partial \Lambda}}_{\mathbb J; \Lambda}}
\end{equation}
where 
$H_{\mathbb J}(\sigma_{\Lambda}\mid \sigma_{\partial \Lambda})$
denotes the object in Eq.(\ref{FH}) restricted to $i$ and $j$ in $\Lambda \cup \partial\Lambda$, $\sigma_{\lambda}$ and $\sigma_{\partial \lambda}$ are notations for the collections of spins on these respective sites and the spins 
$\sigma_{\partial \lambda}$ are to be considered as fixed, $Z^{\sigma_{\partial \Lambda}}_{\mathbb J; \Lambda}$ is the partition function which normalizes $\mathbb{P}$.  
Of exclusive interest in this work are the {\it one} ([1]) and {\it free} ([f]) boundary conditions.  The former case is defined by simply setting all of $\sigma_{\partial \lambda}$ to be in the first spin state.  (Since the Hamiltonian is completely symmetric under permutations of spin--states, all results concerning [1]--boundary conditions apply equally well to the 
$(q-1)$ others that are defined similarly.)
The latter is described by the stipulation that 
$J_{i,j}$ is set to zero whenever $i \in \Lambda$ and $j \in \partial \Lambda$.  Infinite volume thermodynamics is defined via the partition function 
in the usual way.  All the standard thermodynamic functions (and their derivatives with respect to external fields) emerge as almost sure quantities which are independent of boundary conditions; c.f.~\cite {GvE} and references therein.

For finite volume, the quenched measures are defined by averaging the probabilities in Eq.(\ref{Gibbs weights}) over all realizations of the interactions according to $\prod_{\langle i,j \rangle}dF(J_{i,j})$ (written as $dF$) -- where the product runs over the appropriate set, usually $i$ and $j$ in $\Lambda\cup\partial\Lambda$).  These averaged objects will be referred to as the finite volume {\it quenched} measures and denoted by $\mathbb Q_{F;\Lambda}^{\sharp}$ with the superscript denoting one of the two types of boundary conditions described.  
As far as infinite volume limits are concerned, there are two alternative scenarios:  (1) for fixed $\mathbb J$ take $\Lambda \nearrow \mathbb Z^2$ and (somehow) average the resultant measure over $\mathbb J$.  (2)  Take the (vague) infinite volume limit of the finite volume quenched measures described above.  
It turns out that for the pertinent cases at hand, the order of the procedures is immaterial and the result is independent of how 
$\Lambda \nearrow \mathbb Z^2$.  We will provide a brief sketch of how this is established in the appendix.

Below we summarize our claims concerning these quenched measures.

\noindent {\bf Proposition A1}  {\it Consider the Hamiltonian in Eq.(\ref{FH}).  For all F satisfying the ferromagnetic condition, that is $J_{i,j} \geq  0$ with probability one,
the limiting measures}
\begin{equation}
\mathbb Q^{[1]}_{F}  
=  \lim_{\Lambda \nearrow \mathbb Z^2} \mathbb Q^{[1]}_{F; \Lambda}
\end{equation}
{\it exists (independently of how $\Lambda \nearrow \mathbb Z^2$) and similarly for $\mathbb Q^{[\text{f}\thinspace]}_{F}$.  Furthermore, the measures}
$\mathbb Q^{[1]}_{F}$
{\it and
$\mathbb Q^{[\text{f}\thinspace]}_{F}$
 are invariant under $\mathbb Z^2$ shifts -- as well as other $\mathbb Z^2$ symmetries.}
\medskip

\noindent {\bf Proposition A2}  
{\it Consider the Hamiltonian in Eq.(\ref{FH}).  If}
$\mu_{\mathbb J}^{[1]}$ and 
$\mu_{\mathbb J}^{[\text{f}\thinspace]}$ {\it are limiting Gibbs measures (which also exist independently of how the infinite volume limit is taken) then the average of these measures is well--defined and equal to their respective $\mathbb Q$ counterpart.}
{\it Finally, if F is such that the spontaneous magnetization as defined thermodynamically (or as will be discussed subsequent to Eq.(\ref{MP})) is zero then the limiting quenched measure -- satisfying all properties of Proposition A1 above and A3 below -- is unique.} 
\medskip

\noindent {\bf Proposition A3}
{\it The limiting measure}
$\mathbb Q^{[1]}_{F}$
{\it satisfies the strong mixing condition.  So, in particular, if the magnetization vanishes the unique measure is strongly mixing.}
\medskip

While most likely these are non--Gibbsian measures, as is evidenced by the result in \cite{EMSS} on a related system, they are physically motivated
and, possibly, experimentally accessible. Notwithstanding, at self--dual points (and presumably at other critical points) the combined efforts of species 2 through $q$ will fail to achieve percolation in spite of their high density.  Furthermore in the dilute case (and related cases) despite a low density, the type--1 spins can, on their own, achieve percolation in the $\mathbb Q^{[1]}_{F}$ -- measures.

\subsection{\large{Statement of Main Results}}
For  $0 < J < \infty$, the dual coupling is defined by
\begin{equation}
\label{dual}
J^{\ast}(J)= \log\left(1+\frac{q}{e^{J}-1}\right).
\end{equation}
and the dual model is defined by the assignment of the coupling $J^{\ast}(J_{i,j}) \equiv J^{\ast}_{i,j}$ to the bond 
$\langle i^{\ast},j^{\ast} \rangle$ of the dual lattice $(\mathbb Z + \frac 12)^{2}$ that is traversal to the bond $\langle i,j \rangle$.
A model is self--dual \negthinspace
\footnote{For models that do not respect all of the $\mathbb Z^2$ symmetries, a more general definition of self--duality is possible e.g.~the vertical bonds distributed as the dual of the horizontal bonds and vice versa.  While most of our results go through easily in these cases, we make no further reference to these extensions since the present purpose is to construct ``natural measures'' on $\mathbb Z^2$}
if
\begin{equation}
\label{self dual}
dF(J_{i,j})  =  dF(J^{\ast}_{i,j}).
\end{equation}
Our result on the bond--random Potts ferromagnets reads as follows:
\begin{theorem}
\label{main theorem}
Suppose that for the ferromagnetic bond strength distribution $F$, in both the direct and the dual model, the spontaneous magnetization vanishes.  Then, with probability one, in the (unique) limiting quenched measure there are infinitely many circuits surrounding the origin such that the spin--type is constant throughout the circuit and, moreover, there are infinitely many such circuits of each spin--type.  In particular, there is no percolation (or even $\ast$--percolation) in the various marginal measures which identify as many as $(q-1)$ of the spin--states as a single state notwithstanding that the density of this amalgamation is
$1-\frac{1}{q}$.
\end{theorem}
\begin{corollary}
\label{KORE}
For a 2D disordered Potts model at a point of self--duality, the (hypotheses and) conclusions of Theorem \ref{main theorem} hold.
\end{corollary}
Our next result concerns situations which {\it do} have percolation:
\begin{theorem}
\label{dilute}
Consider the dilute model with parameters $a$ and $\beta$ defined by
$J_{i,j} = \beta$ with  probability $a$ and zero otherwise.
If $a > \frac 12$ there is a $\beta_c < \infty$ such that for all $\beta > \beta_c$, the magnetization  $m(\beta,a)$, given for [1] boundary conditions by 
the excess density of species $1$ above $\frac 1q$, satisfies
the inequalities
\begin{equation}
0 < m(\beta,a) < P_{\infty}(a)
\end{equation}
where  $P_{\infty}(a)$ is the percolation probability in the independent bond--model on 
$\mathbb Z^2$.  In particular, for all $\beta > \beta_c$, in the measure $Q^{[1]}_{F}$, there is (with probability one) an infinite cluster of species 1 while the overall density of this species is $m(\beta,a) + \frac 1q \leq P_{\infty}(a) + \frac 1q$; by considering large $q$ and $a$ close to $\frac 12$, this density can be made as small as desired.
\end{theorem}

Both features may be exhibited in a single model:  
\begin{theorem}
\label{both}
Let $a \in (0,1)$ and $K$ denote a bounded random variable satisfying 
$K \geq b$ with probability one with $b$ considered ``large''.
Consider a disordered Potts ferromagnet with strong and weak bonds:
Suppose that for each 
$\langle i,j \rangle$, 
with probability $a$, $J_{i,j}$ is equal, in law, to $K$ while otherwise, with probability $(1-a)$, it is equal in law to $J^{\ast}(K)$.  
Then for $b$ large, the magnetization is very nearly $P_{\infty}(a)$ and the overall density in the measure  
$Q^{[1]}_{F}$ is very nearly $P_{\infty}(a) + \frac 1q$.  Since the model is self--dual at $a = \frac 12$, both the high and low density percolative phenomena occur as the parameter $a$ passes through $\frac 12$.
\end{theorem}

\subsection{\large{Graphical Representations, Dual Models and Dominations}}

For $\langle i,j \rangle$ a neighboring pair let us define
\begin{equation}
\label{R}
R_{i,j} = e^{J_{i,j}} - 1.
\end{equation}
As is well known \cite{FK} the model admits the {\it random cluster} representation:  In finite volume, if $\omega$ is a configuration of occupied and vacant bonds (or edges) 
the probability of $\omega$ is given by
\begin{equation}
\mathbf P_{\mathbb J;\Lambda}^{\sharp}(\omega) \propto \prod_{\langle i,j \rangle \in \omega}
R_{i,j}q^{c^{\sharp}(\omega)}
\end{equation}
where $\langle i,j \rangle \in \omega$ represents the event that the particular bond is occupied and $c^{\sharp}(\omega)$ denotes the number of connected components which are counted by rules ($\sharp$) in accord with conditions specified at the boundary: $\sharp = [1]$ and 
$\sharp = [\text{f}\thinspace]$.  In the former case -- sometimes called the {\it wired} measures -- $c^{[1]}(\omega)$ counts all clusters that are attached to the boundary as part of the same connected component while $c^{ [\text{f}\thinspace]}(\omega)$ simply counts the number of components in $\omega$ by the conventional definition. Given these random cluster measures, it is possible to write down the conditional probabilities of spin configurations given a bond configuration. This is done by insisting that the spin--value is constant throughout each component of $\omega$ and, except for the components attached to the boundary, assigning the spin--types to each component independently and with equal probability.  As for the boundary component, in so far as concerns the two setups in this work, the procedures are simple:  For  $[\text{f}\thinspace]$, nothing special is done -- all boundary components are treated like the internal components.  For the wired or [1] case, all components of the boundary are set to the first spin--state.  
Thus, for example, the finite volume magnetization at the origin for specified $\mathbb J$ is given by 
\begin{equation}
\label{MP}
m_{\mathbb J; \Lambda}(0) \equiv 
\mathbb E^{[1]}_{\mathbb J;\Lambda}(\delta_{\sigma_0 ,1}) - \frac{1}{q}  =  
\mathbf P^{[1]}_{\mathbb J ;\Lambda}(\{0 \leftrightarrow \partial \Lambda \})
\end{equation}
where $\{0 \leftrightarrow \partial \Lambda \}$ is the event that the origin is connected to $\partial \Lambda$ by an occupied path.  The average magnetization at the origin
$\bar{m}_{\Lambda}^{[1]}(0)$ is obtained by averaging (\ref{MP}) over $F$. The infinite volume spontaneous magnetization $\bar{m}^{[1]}(0)$   is obtained by taking the limit $\Lambda  \nearrow \mathbb Z^d$ of $\bar{m}_{\Lambda}^{[1]}(0)$. We call $\bar{m}^{[1]}(0)$ the spontaneous magnetization of the system.

To conclude:  on the basis of Eq.(\ref{MP}) the magnetization at the origin is non--zero in any given quench (realization) if and only if the right hand side does not tend to zero as $\Lambda \nearrow \mathbb Z^2$; i.e.~percolation and spontaneous magnetization are synonymous.
That the limiting percolation density exists (with or without the quenched average) is well known and anyway can be derived on the basis of what is discussed later.  The fact that percolation probability is equal to the thermodynamically defined spontaneous magnetization has been proved elsewhere, see \cite{GvE} and references therein, as well as \cite{ACCN} and \cite{ACCN2}.

In finite volume, the dual model is defined as follows:  If $\langle i,j \rangle$ is an edge of $\mathbb Z^2$, let $\langle i^{\ast},j^{\ast} \rangle$ denote the corresponding edge of $(\mathbb Z+\frac 12)^{2}$ and let
\begin{equation}
R^{\ast}_{i^{\ast} \negthinspace\negthinspace ,j^{\ast}} = \frac{q}{R_{i,j}}.
\end{equation}
which is the equivalent to Eq.(\ref{dual}) via Eq.(\ref{R}).  
For a finite $\Lambda \subset \mathbb Z^2$ -- with $\Lambda$ regarded as a graph -- consider the dual graph, $\Lambda^{\ast}$ consisting of all (dual) edges corresponding to the (direct) edges in $\Lambda$ and the collection of (dual) sites which are the endpoints of these dual edges.  As is well known, the model on the dual graph with parameters 
$R^{\ast}_{i^{\ast}\negthickspace,j^{\ast}}$ has configurations which are in one--to--one correspondence with (and have the same probabilities as) the original setup on $\Lambda$.  Of course, some attention must be paid to the conditions at the boundary.  All that is needed in this work is the readily verified 
fact that the model with wired boundary conditions on $\Lambda$ associates with the model with free boundary conditions on $\Lambda^{\ast}$ and vice versa. We will refer to the initial model as the {\it direct} model and the induced distribution for the 
$J^{\ast}_{i^{\ast}\negthinspace, j^{\ast}}$ -- collectively denoted by $\mathbb J^{\ast}$ -- by $F^{\ast}$.  Of course when it comes to integration, we may use the ``direct'' $dF$.

A model is said to be {\it self--dual} if the probability distribution of the $\{J^{\ast}_{i,j} \}$ is the same as that of original,  e.g.~Eq.(\ref{self dual}), or equivalently, 
\begin{equation}
R_{i,j}  =_{d}  \frac{q}{R_{i,j}}.
\end{equation}
The ``nicest'' examples concern a self--duality which is achieved according to a temperature parameter, in which case one can speak of the self-dual temperature $\bar{\beta}$, c.f.~Eq.(30) in \cite{CkS}, but this will not be necessary in the present work.

For the sake of completeness, let us recapitulate in brief (special cases of) the domination arguments that were derived in \cite{ACCN, ACCN2}.  For a fixed bond $\langle i,j \rangle$ in finite volume with free or wired boundary conditions, let us calculate the conditional probability that the bond is occupied.
It is not hard to see that this is exactly
\begin{equation}
\label{D1}
p_{i,j}  =  \frac{R_{i,j}}{1 + R_{i,j}}
\end{equation}
if $\{i \leftrightarrow j \}$ while the probability is 
\begin{equation}
\label{D2}
p_{i,j}^{\text{eff}} = \frac{R_{i,j}}{q + R_{i,j}} = \frac{p_{i,j}}{p_{i,j} + q(1-p_{i,j})}
\end{equation}
if the endpoints are not connected.  (In the former, the number of components is unaffected while in the latter, it is reduced by one.)  
It therefore follows from elementary considerations that for each quench, the random cluster measures -- wired or free -- dominate the independent bond measures at parameters $p^{\text{eff}}_{i,j}$ and are dominated by independent bond measures with parameters $p_{i,j}$.  The ease of calculating quenched averages of independent bond measures (percolation on percolation) is what lead to the asymptotically sharp results of \cite{ACCN2}.  Indeed, as is already seen, when the relevant $R$--parameter is large compared with $q$, the upper and lower estimates do not differ by much.  

\subsection{\large{Proofs of Main Results}}

We start with a preliminary result which is most of what is needed for the proof of Theorem \ref{main theorem}.
\begin{proposition}
\label{proposition}
Let $F$ denote a distribution of couplings such that the spontaneous magnetization of the dual model is zero.  Then with probability one for both 
$\mathbb Q^{[1]}_{F}$
and
$\mathbb Q^{[\text{f}\thinspace]}_{F}$
there are infinitely many circuits surrounding the origin such that, within each circuit, the spin--type is constant.
\end{proposition}
\begin{proof}
Let $\epsilon > 0$ and let $\tilde V \subset \mathbb Z^2$ denote a finite set containing the origin.  Let $\tilde \Lambda \supset \tilde V$ and $\mathcal D_{\tilde V \negthickspace,  \tilde \Lambda}$ denote the event
\begin{equation}
\mathcal D_{\tilde V \negthickspace,  \tilde \Lambda}  =  
\{\sigma \mid \exists  \text{ a circuit of constant spin--type separating } \partial \tilde V 
\text{ and } \partial \tilde \Lambda\}
\end{equation}
where by circuit it is meant a graph theoretical ``cycle'' of vertices.
Then, as we shall see, it is sufficient to prove (for arbitrary fixed $\tilde V$ and $\epsilon$) that 
$\mathbb Q^{[\text{f}\thinspace]}_{F}
(\mathcal D_{\tilde V \negthickspace,  \tilde \Lambda})
> 1 - \epsilon$ whenever $\tilde \Lambda$ is sufficiently large, and similarly for 
$\mathbb Q^{[1]}_{F}$.  We let $V$ -- and $\Lambda$ -- denote sets similar to their tilde counterparts enhanced by a layer or two at the boundary to avoid discussion of inconsequential provisos caused by discrete lattice effects.   Let $\Lambda^{\ast}$ and $V^{\ast}$ denote the dual sets and let 
$\Upsilon = \frac{\epsilon}{|\partial V^{\ast}|}$.  
Since the quenched magnetization in the dual measure vanishes by hypothesis, for every $i^{\ast} \in \partial V^{\ast}$ the average magnetization at $i^{\ast}$ in the one--boundary conditions in $\Lambda^{\ast}$ (for the dual model) tends to zero as $\Lambda^{\ast}$ gets large.  Thus, in a large enough volume, for all such $i^{\ast}$,
\begin{equation}
\overline m_{F^{\ast};\Lambda^{\ast}}(i^{\ast}) \equiv
\int dF \thinspace \mathbb E^{[1]}_{\mathbb J^{\ast};\Lambda^{\ast}}(\sigma_{i^{\ast}} - \frac{1}{q})
< \Upsilon.
\end{equation}
However, according to Eq.(\ref{MP}), the integrand is the probability, in the dual model, of a dual connection between $i^{\ast}$ and $\partial \Lambda^{\ast}$.  Thus
\begin{equation}
\label{here}
\int dF\thinspace \mathbf P^{[1]}_{\mathbb J^{\ast};\Lambda^{\ast}}
(\{\partial V^{\ast} \leftrightarrow \partial \Lambda^{\ast}\})
\leq \int dF \left[\thinspace
\sum_{i^{\ast} \in \partial V^{\ast}}
\mathbf P^{[1]}_{\mathbb J^{\ast};\Lambda^{\ast}}
(\{i^{\ast} \leftrightarrow \partial \Lambda^{\ast}\})
\right]
\leq |\partial V^{\ast}|\Upsilon \leq \epsilon.
\end{equation}
However, $1 - \mathbf P^{[1]}_{\mathbb J^{\ast};\Lambda^{\ast}}
(\{\partial V^{\ast} \leftrightarrow \partial \Lambda^{\ast}\})$ is the probability in the direct model 
of an occupied circuit separating $\partial V$ from $\partial \Lambda$ in the transformed boundary conditions.  So we may write
\begin{equation}
\label{integrand}
\int dF \thinspace \mathbf P^{[f]}_{\mathbb J; \Lambda}(\{ \exists \text{ occupied circuit separating }
\partial V \text{ and } \partial \Lambda \}) \geq 1- \epsilon.
\end{equation}
But each spin realization associated with such a random cluster event has a circuit of the stated type and we have obtained the desired result in finite volume with free boundary conditions:
\begin{equation}
\label{there}
\mathbb Q^{[\text{f}\thinspace]}_{F;\Lambda}(\mathcal D_{\tilde V \negthickspace,  \tilde \Lambda})
\geq 1 - \epsilon.
\end{equation}
The result for infinite volume follows by monotonicity of the integrand in Eq.~(\ref{integrand}) in the system size:  As discussed in the appendix, for free boundary conditions, the integrand increases when we consider the same event (involving $\partial \Lambda$) in a volume $\Lambda^{\prime} \supset \Lambda$.  For wired boundary conditions, the result follows because the in any volume, the integrand increases if we replace [f] by [1].

Now, to see that the above implies the presence of {\it infinitely} many circuits,  let $(\tilde {\Lambda}_k \tilde {V}_k )$ denote a sequence of shapes as above (which exhaust the lattice) and satisfy 
$\tilde {\Lambda}_{k-1}\subset \tilde {V}_k \subset \dots$ such that the probability of a circuit in the $k^{\text{th}}$ pair exceeds $1-\epsilon_{k}$ with $\sum_k \epsilon_k < \infty$.  By the Borel--Cantelli lemma, with probability one, only a finite number of the prescribed circuits fail to appear. 
\end{proof}
\noindent{\it Proof of Theorem \ref{main theorem} }  
Since the dual magnetization vanishes we already have, according to the previous proposition, the part concerning the infinitely many circuits.  What is lacking is a proof that these circuits are not not all of the same spin--type (as would indeed be the case if the spontaneous magnetization were positive).   Now, using the fact that the direct magnetization also vanishes, we may demonstrate, along the lines of 
Equations (\ref{here}) through (\ref{there}) that outside any finite $V$ but inside 
a sufficiently large $\Lambda$ the average probability of observing the dual of a circuit composed of {\it vacant} bonds is close to one -- even with (direct) wired boundary conditions.

The upshot, when both magnetizations vanish, is that even in finite volume, with high probability many circuits of dual bonds and many circuits of direct bonds surround the given $V$.  Since both the direct and dual circuits can be ``freshly constructed'' at an increasing sequence of scales, it follows that some portion of these circuits separate each other.  We are now, more or less, in the same situation as Theorem \ref{DaC1} for the DaC model:  There are many direct rings separated from each other (and the boundary) by the dual rings.    These direct rings are therefore ``at liberty'' to take on any of the $q$  spin--values.  In particular for any $s \in \{1, \dots q\}$ and any finite $V$,
\begin{equation}
\mathbb Q^{[1]}_{F;\Lambda}
(\{\exists \text{ a circuit of spin--type } s \text{ surrounding } V \}) \longrightarrow 1
\end{equation}
as $\Lambda \nearrow \mathbb Z^2$.  The result also holds in $\mathbb Q^{[\text{f}\thinspace]}_{F;\Lambda}$ (which anyway leads to the same limiting measure as we show in Proposition A2).
The conclusion is that with probability one, in the limiting measure there are infinitely many circuits of all types surrounding the origin.  This establishes the first claim.  It also establishes the absence of percolation for any amalgamation--alliance since an infinite path (connected or $\ast$--connected) of the alliance without the $s^{\text{th}}$spin--state, starting from the origin, is prevented by any of the $s^{\text{th}}$ state's circuits, the presence of which has probability one.  Finally, as is evident from the absence of magnetization, the population density of all species is exactly $\frac{1}{q}$. \qed

\noindent{\it Proof of Corollary \ref{KORE}} \thickspace
As was demonstrated in \cite{CkS}, Theorem 1$^{\prime}$, the magnetization of the disordered 2D Potts models vanishes at self dual points.  This applies equally well to the dual model. 
\qed

\noindent{\it Proof of Theorem \ref{dilute}} \thickspace
This result is, to the largest extent, contained in \cite{ACCN2} so we will be succinct.  In a given quench, each edge corresponding to a zero coupling bond adds nothing to the ferromagnetism.  The active bonds must be ``reoccupied'' (i.e. occupied in the random cluster problem on this media) with a conditional probability bounded above by $R/[1 + R]$ and below by $R/[q + R]$ where $R = e^{\beta} -1$.  The fraction of these reoccupied bonds that belong to an infinite cluster (in the limiting [1]--state) constitute the magnetized fraction.  After performing a quenched average, it is seen that the magnetization is positive whenever 
\begin{equation}
a\frac{R}{q + R} > \frac12
\end{equation}
but is bounded above by $P_{\infty}(\frac{aR}{q+R}) < P_{\infty}(a)$.  This is small and positive for all $\beta$ large enough and $a$ close to $\frac 12$. The density of ones, in the one state is just this magnetization plus the ambient  $\frac 1q$.  \qed
\medskip

\noindent{\it Proof of Theorem \ref{both}} \thickspace  For simplicity we will treat just the case where with probability $a$,  $R = e^{J} - 1 = e^{b} - 1 \equiv B$  and with probability $(1-a)$, $R = q/B$.  Obviously, for $a = \frac 12$, the model is self--dual.  To prove the remaining statements, we reiterate that
\begin{equation}
P_{\infty}(p^{\text{eff}})  \leq  m(B,a)  \leq P_{\infty}(p)
\end{equation}
where, in general, $p^{\text{eff}} = \int dF[R/(q + R)]$, etc.  Thus it is sufficient to find $B$ and $a$ such that both $p^{\text{eff}}$ and $p$ are in slight excess of $\frac 12$.   In this simple case, we may write 
\begin{equation}
p^{\text{eff}}  =  a\frac{B}{B+q}  + (1-a)\frac{1}{1 + B}
\end{equation}
and 
\begin{equation}
p  =  a\frac{B}{1 + B}  +  (1-a)\frac{q}{B + q}.
\end{equation}
It is readily seen that when $a \gtrsim \frac 12$ and $B \gg q$ the desired conditions are met.
\qed
\section{Discussion}

We have shown here that for the disordered ferromagnetic $q$-state Potts model on $\mathbb Z^2$ at points of self--duality there is no percolation for any amalgamation of $q-1$ components in the unique translation invariant mixing measure $\mathbb Q_{F}$. This transpires despite the  fact that the density of these combined $q-1$ states is $1-q^{-1}$ which will be arbitrary close to 1 as $q \rightarrow \infty$. Of course we expect that the above holds at all critical points of these models.  In particular, we suppose, on general grounds, that the conditions of Theorem \ref{main theorem} (namely no percolation and no dual percolation in the graphical representation) holds at any critical point in a model of this sort.  

It is further expected that as soon as the temperature is decreased below the critical temperature or the overall interaction strength increased, the magnetization will rise continuously from zero.  
If this scenario is correct,  i.e. if the magnetization (like the energy density -- which in this context may be taken to mean the bond density) is continuous for this system, it would follow that in each of these states we would have percolation at a small density. 

Unfortunately, even aided with self--duality, we cannot make such an assertion.  There are however some physical arguments
supported by simulations which indicate that this is indeed the case, see ~\cite{jacobsen,cardy} and references therein. 
We may use duality to prove the weaker assertion that the magnetization is not already positive at a self--dual point (c.f.~the discussion in \cite{BaC}).  Indeed, this would imply the identical circumstance in the dual model -- with appropriately modified boundary conditions -- and the latter would be a non--percolative state from the perspective of the direct model which would mean the existence of two states with differing energy density which is ruled out by the result of \cite{AW}.  More succinctly, there cannot be a point where the model exhibits a percolative and a non---percolative state -- so the magnetization indeed vanishes at a self--dual point.  But this does not quite rule out the possibility of an interval of critical behavior -- with vanishing magnetization -- 
surrounding a self--dual point with a discontinuity in the magnetism at the endpoint and {\it no} coexisting non--percolative state at the point of discontinuity.  

Trivial examples of discontinuous order parameters coinciding with critical transitions are abundant in short--range 1D systems at zero temperature e.g.~the Ising model.  The energy is continuous as $T \downarrow 0$ but $m(0)=1$.  More intricate examples can be found, here we mention two.  First there is the (reinterpretation of the) mean--field $k$--core transition \cite{SLC} where, in the presence of two distinct divergent length scales -- and susceptibilities -- there is a discontinuity in the order parameter.  Second, we mention the well known Thouless effect \cite{T} for the 1D Ising model with ferromagnetic pair--interactions that decay like the inverse square of the separation.  Here, as was proved in \cite{ACCN}, the magnetization is discontinuous at the transition point but this point is critical in the sense of a divergent length scale, a divergent susceptibility and, it is presumed, a continuous energy density.  It is interesting to note that, from the low temperature side, this transition is indeed the endpoint of a critical phase \cite{Chak, IN}.

In this note we have circumvented these mathematical intricacies and demonstrated percolation at small densities by considering dilute and ``nearly dilute'' models.  In the former case, the magnetization is always small and in the latter case, it is demonstratively small in the vicinity of the critical point.  This certainly does not settle the issue of magnetic/percolative continuity in these models, but it does, perhaps, bring us a small step closer.   

\section{Appendix}
{\it Proof of Proposition A1 }  The first crucial ingredient in what is to follow is monotonicity of the associated (fixed coupling) random cluster measures in finite volume.  In particular, if $\Lambda_{1} \supset \Lambda_{2}$ and we compare the wired measure on $\Lambda_{1}$ with (the restriction of) the wired measure in $\Lambda_{2}$ -- with the same $J_{i,j}$'s in the common territory -- then the smaller system FKG--dominates (meaning that both FKG measures satisfy the stated stochastic domination).   For free measures the situation is similar only the domination goes the other way.  The second ingredient is that expectations of spin--functions may be expressed as expectations of random cluster functions which may be easily decomposed into increasing and decreasing functions.  For example if 
$A_{1}\subset \mathbb Z^{2}$ denotes a finite set and $\mathbf 1_{A_{1}}$ denotes the indicator of the event that all spins on $A_{1}$ are of type one, it is not hard to see \cite{ACCN} that 
\begin{equation}
\label{KK}
\mathbb E^{\text{[f]}}_{\mathbb J;\Lambda}(\mathbf 1_{A_{1}}) =  
\mathbf E^{\text{[f]}}_{\mathbb J;\Lambda}
(q^{-K(A_{1})})
\end{equation}
where $K(A)$ is the number of distinct components that intersect $A$.  Observing that $K(A_{1})$ is a decreasing function (and that $1/q$
is less than one) the right hand side of Eq.~(\ref{KK}) is the expectation of an increasing function.  Notice then that monotonicity immediately implies the existence of a limiting probability for this particular cylinder, independently of how the $\Lambda$'s go to $\mathbb Z^{2}$.  

For multiple sets, the situation is only slightly more complicated.  E.g.~consider
$\mathbf 1_{A_{1}}\mathbf 1_{A_{2}}$ (where $\mathbf  1_{A_{2}}$ is the indicator of the event that all spins in $A_2$ are of type {\it two}).  Here,  the product of the two appropriate random cluster functions is almost the right answer except we must insist that there is no connection between $A_{1}$ and $A_{2}$ since the latter would demand that some sites in $A_{1}$ would take the same spin--value as sites in $A_{2}$.  However, this constraint may be written as 
$1 - \mathbf 1_{\{ A_{1} \leftrightarrow A_{2}\}}$ and, we are again in a situation where we may analyze (sums and differences of) increasing functions.  The general formula for arbitrary cylinder function is seen to be
\begin{equation}
\label{product}
\mathbb E^{\text{[f]}}_{\mathbb J;\Lambda}
(\prod_{j=1}^{q}\mathbf 1_{A_{j}})
=
\mathbf E^{\text{[f]}}_{\mathbb J;\Lambda}
(q^{-(\sum_{j}K(A_{j})}\prod_{j \neq k}[1 - \mathbf 1_{\{ A_{j} \leftrightarrow A_{k}\}}])
\end{equation}
and we may use monotonicity in volume to demonstrate the existence of a limiting free measure independent of how the limit is taken.  

For the [1] boundary conditions, the overall situation is similar with just a couple of modifications that need to be made.  The first difference is that now the sets $A_{2}, \dots A_{q}$ must not be allowed a connection to the boundary hence the function in Eq.(\ref{product}) is be modified by the insertion of the product
$\prod_{j>1}[1 - \mathbf 1_{\{ A_{j} \leftrightarrow \partial \Lambda\}}]$.
The second adjustment concerns the interpretation and counting of the number of clusters intersecting the set $A_1$.  Clearly any sites in $A_1$ that are connected to the boundary are ``already'' in the 1--state and, needless to say are considered part of the same cluster.  Thus, $q^{-K(A_1)}$ must be replaced by 
$q^{-K_{W}(A_1)}[{\bf 1}_{\{A_1 \leftrightarrow \partial \Lambda\}^c} + 
q{\bf 1}_{\{A_1 \leftrightarrow \partial \Lambda\}}]$ where $K_W$ denotes the number of clusters counted according to the wired rules.     Note that the second factor may be rewritten as $[1 + (q-1){\bf 1}_{\{A_1 \leftrightarrow \partial \Lambda\}}]$ which, along with 
$q^{-K_{W}(A_1)}$, is manifestly increasing.  Moreover, 
the (increasing) connectivity events $\{ A_{j} \leftrightarrow \partial \Lambda\}$ have the property that if $\Lambda \supset \Lambda^\prime$ then 
$\{ A_{j} \leftrightarrow \partial \Lambda\} \subset \{ A_{j} \leftrightarrow \partial \Lambda^\prime\}$
which is in accord with the decreasing tendency of the measures in increasing volume.  Thus, a limit emerges again by means of monotonicity properties.  

It is clear that both of the limiting measures are translation invariant -- as well as enjoying all other $\mathbb Z^{2}$--symmetries.  For example if $\mathcal B_{A}$ is a local event which depends only on the spins in $A$ and $\mathbf T$ is some $\mathbb Z^{2}$ translation, it is manefest that 
\begin{equation}
\mathbb Q^{\text{[f]}}_{F,\mathbf T(\Lambda)}(\mathcal B_{\mathbf T(A)}) =
\mathbb Q^{\text{[f]}}_{F,\Lambda}(\mathcal B_{A}).
\end{equation}
However, by the existence of limiting measure (independent of how $\Lambda \nearrow \mathbb Z^{2}$) the left hand side converges to 
$\mathbb Q^{\text{[f]}}_{F}(\mathcal B_{\mathbf T(A)})$
while the right side tends to 
$\mathbb Q^{\text{[f]}}_{F}(\mathcal B_{A})$ which is the desired invariance.  Other invariance properties and these properties for the other measure follow from an identical argument.  The claims of Proposition A1 are all established.
\qed
\medskip

\noindent{\it Proof of Proposition A2 }
We start with a proof of the second statement (uniqueness of the quenched measure if the magnetization vanishes) since we will use these arguments in the subsequent proofs.  We consider a sequence of volumes $(\Lambda)$ with boundary conditions which may depend (measurably, of course) on $\mathbb J$. Now with
$\mathbb J$ fixed, the measure obtained by any particular spin--configuration at the boundary will lead to some form of a random cluster measure.  The details need not concern us here except to say that all such measures (and therefore combinations thereof) are dominated by the wired measure.   
We again consider some local spin--function which we decompose into a sum of monotone random cluster functions.  We let  $V_1 \subset V_2 \subset \Lambda$ with $V_{2}$ large compared to $V_{1}$ and $\Lambda$ much larger still.  Since the magnetization vanishes, 
using the argument of Proposition \ref{proposition} we are assured, even for wired boundary conditions on $\Lambda$, that a circuit of dual bonds separates $\partial V_{1}$ from $\partial V_{2}$ with probability tending to one as $V_{2}$ gets large.  This is a negative 
(decreasing)
event so the statement also holds in the measure coming from the arbitrary boundary condition -- which we now permit to tend to an infinite volume limit in any fashion.
But meanwhile in the vicinity of $V_1$ and $V_2$, as far as the random cluster problem is concerned,
the situation inside the outermost ring separating $\partial V_1$ from $\partial V_2$ is equivalent to free boundary conditions on this ring.  Of course the statistics of the location of such a ring will depend in detail on $\mathbb J$ but given such a ring, the conditional measure inside $V_{1}$ is dominated by the one with free boundary conditions on $V_{2}$ and dominates the one with free boundary conditions on $V_{1}$.  The conclusion is that with large probability the random cluster expectations lie between the free boundary condition problems on $V_{1}$ and $V_{2}$ and we may safely take quenched averages.  By allowing $V_{2}$ and then $V_{1}$ to exhaust $\mathbb Z^{2}$ we have proved any infinite volume measure is the same as the limiting infinite volume free measure.  

To prove the first statement, let us distinguish two cases depending on the vanishing or non--vanishing of the spontaneous magnetization.  When there is no magnetization, the argument is only a slight reworking of the discussion in the previous paragraph.  
First, by appealing to stochastic (FKG) monotonicity in volume, arguments similar to ones already used show that for a.e.~$\mathbb J$, the limiting $\mu_{\mathbb J}$ is unique, independent of how the limit is taken, etc.  (Thus, we are technically proving a stronger result than the one claimed.)  So, in particular, for a.e. $\mathbb J$, the infinite volume measure may as well be considered as the limit of measures constructed in finite volume with free boundary conditions.
Now consider, for fixed $\mathbb J$, the thermal average of some local spin event $\mathcal A$.  As before, this may be expressed as a sum involving expectations of increasing random cluster functions. 
Let $V_1$ denote a large volume which contains the support of the spin function and $V_2$ containing $V_1$ larger still.  With the vanishing of the (quenched) magnetization, again using the arguments 
in the proof of Proposition \ref{proposition}, we are assured that with high probability (in the random--coupling/random cluster {\it joint} measure) there is a dual ring of vacant  random cluster bonds separating the boundaries of the $V$'s; in particular, with probability tending to one as $V_2$ tends to $\mathbb Z^2$.  
This accomplishes two tasks: first it establishes $\mathbb J$--measurability of these thermal expectations and second it shows that their quenched average is close to the finite volume quenched average with various free boundary conditions.  By allowing the volumes to exhaust $\mathbb Z^{2}$ we have established interchangeability of the two limiting procedures. 

	When the magnetization does not vanish, the argument is somewhat more intricate.  To start off, we again claim that for fixed $\mathbb J$, various infinite volume measures exist.  In particular, here we will be interested in the limiting wired and free random cluster measures, ${\bf P}^{[\text{w}]}_{\mathbb J}$ and ${\bf P}^{[\text{f}]}_{\mathbb J}$ that emerge from sequences of the corresponding finite volume measures.  Furthermore, we will have brief need to refer to the {\it quenched} limiting random cluster measures that arise from these boundary conditions.  
	
		Let $V_k$ denote any standard sequence of volumes tending to $\mathbb Z^{d}$, e.g.~nested squares centered at the origin.  
Let $\Theta_k$ denote the event that there is a ring of dual bonds separating $\partial V_k$ from infinity.  Finally, let ${\bf \Theta}$ denote the event that infinitely many of the $\Theta_k$--events occur.  We claim that if the magnetization is positive, then the set 
$\{\mathbb J\mid {\bf P}^{[\text{w}]}_{\mathbb J}({\bf \Theta}) \neq 0 \}$ has $F$-probability zero.  Indeed, let us first establish that this event is is a tail event.  Foremost, it is manifestly shift--invariant.  Next, supposing that $\mathbb J$ is such that 
${\bf P}^{[\text{w}]}_{\mathbb J}({\bf \Theta}) \neq 0 $, let us demonstrate that 
${\bf P}^{[\text{w}]}_{\mathbb J^{\prime}}({\bf \Theta}) \neq 0 $
if $\mathbb J$ and $\mathbb J^{\prime}$ differ at only a finite number of bonds.  This follows easily since, if $D$ is a finite set of bonds and 
$\mathbb J = \mathbb J^{\prime}$ on $D^{c}$, then for any fixed configuration on $D$, the conditional measures agree.  Therefore, letting $D$ denote the set where $\mathbb J$ and $\mathbb J^{\prime}$ disagree and letting $D_{\emptyset}$ denote the event that all bonds of $D$ are vacant, we have
\begin{equation}
{\bf P}^{[\text{w}]}_{\mathbb J^{\prime}}({\bf \Theta}) \geq 
{\bf P}^{[\text{w}]}_{\mathbb J^{\prime}}(D_{\emptyset})
{\bf P}^{[\text{w}]}_{\mathbb J^{\prime}}({\bf \Theta}|D_{\emptyset}) = 
{\bf P}^{[\text{w}]}_{\mathbb J^{\prime}}(D_{\emptyset})
{\bf P}^{[\text{w}]}_{\mathbb J}({\bf \Theta}|D_{\emptyset})
\geq {\bf P}^{[\text{w}]}_{\mathbb J^{\prime}}(D_{\emptyset})
{\bf P}^{[\text{w}]}_{\mathbb J}({\bf \Theta})
\end{equation}
where the last inequality is by the FKG property.  Noting that by Eqs.(\ref{D1}) and (\ref{D2}), ${\bf P}^{[\text{w}]}_{\mathbb J^{\prime}}(D_{\emptyset})$ is not zero (even in the dilute case) the claim follows.  
So now we must rule out the possibility that 
$\{\mathbb J\mid {\bf P}^{[\text{w}]}_{\mathbb J}({\bf \Theta}) \neq 0 \}$ has $F$-probability one.  Assuming that this is the case, let us demonstrate the implication that there is no percolation in the random cluster version of the quenched free measures.  
To this end, let ${\bf \Theta}^{\ell}$ denote the event that a separating ring occurs {\it inside} $V_\ell$.  Under our assumption we have, for almost every $\mathbb J$, 
\begin{equation}
\lim_{\ell \to \infty}
{\bf P}^{[\text{f}]}_{\mathbb J}({\bf \Theta}^{\ell}|{\bf \Theta}) \geq 
{\bf P}^{[\text{w}]}_{\mathbb J}({\bf \Theta}^{\ell}|{\bf \Theta})  = 1
\end{equation}
Letting $\epsilon \in \mathbb R^{+}$ there is therefore a (random) $\ell$ such that 
\begin{equation}
{\bf P}^{[\text{f}]}_{\mathbb J}({\bf \Theta}^{\ell}|{\bf \Theta}) \geq 1-\epsilon
\end{equation}
and while this $\ell$ may have slow decay, it is clear that it is not defective.  Thus, for any small $\eta$, we may define an $L_{0}$ such that $\ell < L_{0}$ with $F$--probability in excess of $1 - \eta$.  However, it is clear that for every $\mathbb J$ and any $L$ and $L^{\prime}$ with $L^{\prime} > L$, 
\begin{equation}
{\bf P}^{[\text{f}]}_{\mathbb J}({\bf \Theta}^{L}|{\bf \Theta})
=
{\bf P}^{[\text{f}]}_{\mathbb J}({\bf \Theta}^{L}|{\bf \Theta}_{L^{\prime}})
\leq 
{\bf P}^{[\text{f}]}_{\mathbb J, \Lambda_{L^{\prime}}}({\bf \Theta}^{L})
\end{equation}
because free boundary conditions enhance any decreasing event inside.
Hence, for any sequence of volumes tending to infinity with free boundary conditions, the quenched probability that the origin is connected to
$\partial \Lambda_{L_{0}}$ is less than $\epsilon + \eta$; i.e.~there is no percolation in the limiting version of these measures.  
	Meanwhile, there {\it is} percolation in the wired rendition of this measure since, by hypothesis, the magnetization is positive.  {\it A priori} the two measures are stochastically ordered and so, evidently, the ordering is strict.  By (the corollary to) Strassen's Theorem, \cite{Strassen}  -- c.f.~the discussion in \cite{Lig} on page 75 -- this implies that the one--dimensional marginals, which here amounts to the translation invariant FK--bond density, differ strictly.  
	Back in the spin--system, by an argument to be found in \cite{MCLSC} -- presumably known to others -- an exact formula can be presented that relates the bond density in the random cluster models to the energy density in the corresponding spin--system.  The implication is a difference among states, (and hence a discontinuity) in the energy density.  This is forbidden by the result of \cite{AW} where it was shown that for the systems described by the interaction in Eq.(\ref{FH}), the energy density is continuous.  Thus, henceforth, we may safely assume that with $F$--probability one, 
${\bf P}^{[\text{w}]}_{\mathbb J}({\bf \Theta})= 0$.

	Now consider a local spin function which, as before, may be expressed as a sum of terms involving increasing random cluster functions.  Notice that in this case, for the modification of Eq.(\ref{product}) discussed shortly after the display, the condition that certain sets must not be connected to the boundary are now replaced by the condition that these sets are not connected to the infinite cluster which is here identified as a cluster of spin--type 1. 
We may assume the event ${\bf \Theta}^{c}$ so that if $V_1$ (which contains the support of the spin function) is large, we are assured that $\partial V_1$ is connected to this infinite cluster with probability close to one.  Further, if $V_2$ which contains $V_1$ is sufficiently larger still, we may use the absence of dual percolation to generate, with high probability, an additional circuit of occupied bonds separating $\partial V_1$ from $\partial V_2$.  By courtesy of the infinite path emanating from $\partial V_1$, any such ring represents, from the perspective of the spin--system, sites which are a constant spin--value of 1.  Thus, by consideration of the measure conditioned on the outermost separating ring, we have that with high probability, the random cluster functions have expectations which are bounded between expectations in the wired measures on $\partial V_1$ and the wired measure on $\partial V_2$.  And, importantly, in both measures, the boundary conditions can be identified with spin boundary conditions of the 1--type.  The remainder of the proof now follows along similar lines as the previous case.  \qed

\medskip

\noindent{\it Proof of Proposition A3 }  Let $f$ and $\ell$ denote two local spin functions which are assumed to be of the form appearing on the left hand side of Eq.(\ref{product}).  We let ${\bf T}$ denote some $\mathbb Z^2$ translation operator and, for notational simplicity, let $g = {\bf T}(\ell)$ denote the translated version of $\ell$.  The sets corresponding to $g$ will be denoted by $B_1, ... B_q$ and we will use $A$'s for the function $f$.  Since, eventually, $|{\bf T}| \to \infty$ it may be safely assumed that these $A$--sets and the $B$--sets are all disjoint.  

	While the functions $f$ and $g$ individually admit an expansion as on the right hand side of Eq.(\ref{product}), the product of $f$ and $g$ does not expand into the product of the expansions.  Indeed, what we obtain is exactly the form of the expression in Eq.(\ref{product}) with the sets $A_j$ replaced by $A_j\cup B_j$.  Because of these -- and additional upcoming -- difficulties, we shall segregate the high and low temperature cases (meaning vanishing or non--vanishing of the magnetization).  We start with the simpler case in which the magnetization vanishes and thence, according to Proposition A2, we might as well assume that have constructed the limiting measure from free boundary conditions.  
	
	Let $V_1$ denote a box that contains all the $A$--sets and, as previously, $V_2$ a much larger box containing $V_1$ with $\partial V_1$ and $\partial V_2$ well--separated.  Similarly let $V_1^\prime$ and $V_2^\prime$ denote the corresponding sets (with the similar characteristics) for the function $g$.  It will be assumed that ${\bf T}$ is large enough so that $V_2$ and $V_2^\prime$ are disjoint.  We let $\mathcal F$ and $\mathcal G$ denote the random cluster expression for the functions $g$ and $f$ -- i.e.~the term inside the expectation on the right side of Eq.(\ref{product}) -- and $\mathcal K$ the expression for the product $fg$.  Consider the event $R$, that a circuit of vacant bonds separates $\partial V_1$ from $\partial V_2$ and similarly for the event $R^\prime$.  It is claimed that should both (or either) of these events occur then $\mathcal K$ indeed factors into $\mathcal F\mathcal G$, e.g.
\begin{equation}
\mathcal K  {\bf 1}_{R\cap R^\prime} =  [\mathcal F][\mathcal G]{\bf 1}_{R\cap R^\prime}.
\end{equation}
Indeed, under this condition, there cannot be a connection between $V_1$ and $V_1^\prime$
so we have $K(A_j\cup B_j)  =  K(A_j) + K(B_j)$ and similarly, the event 
$\{A_i\cup B_i \leftrightarrow A_j\cup B_j\}^c$ reduces to 
$\{A_i\leftrightarrow A_j\}^c\cap\{B_i \leftrightarrow B_j\}^c$.  
	
	The reader is reminded that the absence of magnetization implies that with high probability (with respect to the joint measure) such rings are likely to occur.  If we therefore defer attention from the unlikely event that either of these ring events fail, we are left, for fixed $\mathbb J$, with the random cluster expectation of a sum of the form $\sum_{i,j}\lambda_{i,j}\alpha_i\beta_j$ where the $\alpha$'s and $\beta's$ are non-negative increasing random cluster functions -- bounded by 1 -- and the $\lambda$'s are of undetermined sign.  Let us therefore consider, e.g.~in free boundary conditions on some much larger $\Lambda$ containing all of the $V$'s, the generic expectation 
${\bf E}^{\text{[f]}}_{\mathbb J;\Lambda}[(\alpha)( \beta)({\bf 1}_{R\cap R^\prime})]$.  On the one hand,
\begin{equation}
\label{KAY}
{\bf E}^{\text{[f]}}_{\mathbb J;\Lambda}[(\alpha)( \beta)({\bf 1}_{R\cap R^\prime})]
\geq 
{\bf E}^{\text{[f]}}_{\mathbb J;\Lambda}[\alpha\beta] - 
{\bf P}_{\mathbb J;\Lambda}(\{R\cap R^\prime\}^{c})
\end{equation}
and, in the above mentioned spirit, let us focus only on the first term.  Using FKG, 
\begin{equation}
\label{double 1}
{\bf E}^{\text{[f]}}_{\mathbb J;\Lambda}[\alpha\beta] \geq 
{\bf E}^{\text{[f]}}_{\mathbb J;\Lambda}[\alpha]
{\bf E}^{\text{[f]}}_{\mathbb J;\Lambda}[\beta].
\end{equation}
We must now perform the average over $\mathbb J$ of both sides.  We claim that
\begin{equation}
\label{double 2}
\int dF\ {\bf E}^{\text{[f]}}_{\mathbb J;\Lambda}[\alpha]
{\bf E}^{\text{[f]}}_{\mathbb J;\Lambda}[\beta] \geq 
\int dF\ {\bf E}^{\text{[f]}}_{\mathbb J;\Lambda}[\alpha]
\int dF\ {\bf E}^{\text{[f]}}_{\mathbb J;\Lambda}[\beta].
\end{equation}
Indeed, the quantity 
${\bf E}^{\text{[f]}}_{\mathbb J;\Lambda}[\alpha]$ is the random cluster expectation of an increasing function which as is not hard to check, is an increasing function of $\mathbb J$ and we may again use the FKG inequality because the {\it independent} measure enjoys this property.  

	On the other hand, we may take the expectation on the left side of Eq.(\ref{KAY}) and expand according to which pair of rings is the outermost pair in $V_2\setminus V_1$, and, respectively, $V_2^{\prime}\setminus V_1^{\prime}$.  Under this conditioning, the expectations are independent; both with regards to the fixed $\mathbb J$ random cluster measures, where each of the factors behaves like the free boundary measure at the respective conditioned rings, {\it and} with regards to the $\mathbb J$'s because the regions of interest are disjoint.  Moreover, since $\alpha$ and $\beta$ are increasing functions, the conditional expectations are bounded above by the expectations with free boundary conditions on the boxes $V_2$ and $V_2^{\prime}$.  Thus, all in all,
\begin{equation}
\label{RAY}
{\bf E}^{\text{[f]}}_{\mathbb J;\Lambda}[(\alpha)( \beta)({\bf 1}_{R\cap R^\prime})] \leq 
{\bf P_{\mathbb J; \Lambda}}(\{R\cap R^\prime\})
{\bf E}^{\text{[f]}}_{\mathbb J;V_2}[\alpha]
{\bf E}^{\text{[f]}}_{\mathbb J;V_2^{\prime}}[\beta] \leq 
{\bf E}^{\text{[f]}}_{\mathbb J;V_2}[\alpha]
{\bf E}^{\text{[f]}}_{\mathbb J;V_2^{\prime}}[\beta]
\end{equation}
where the rightmost term stays factored in the quenched average (over $\mathbb J$) due to the aforementioned independence.  
In light of Eq.(\ref{double 2}), Eq.(\ref{RAY}) and their immediate consequences, the individual pieces of the function have, after expectation, essential upper and lower bounds in terms of the product evaluated in infinite volume.  The original functions can then be reconstituted and, as $\Lambda$ and the $V$'s tend to $\mathbb Z^2$ (the latter as the magnitude of the translation becomes large) we obtain the limiting equality
\begin{equation}
\lim_{|{\bf T}|\to\infty}
\mathbb E^{\text{[f]}}_{F}(f\thinspace {\bf T}(\ell)) -  \mathbb E^{\text{[f]}}_{F}(f)
\mathbb E^{\text{[f]}}_{F}(\ell) = 0.
\end{equation} 

The case of the wired/type--[1] boundary conditions follows a similar tack but is somewhat more arduous.  The setup is pretty much the same as in the previous paragraphs:  $|{\bf T}|$ is large, $V_1$ and $V_1^\prime$ are large volumes which contain the supports of the two functions and are, in turn, each contained in $V_2$ and $V_2^\prime$ which themselves are disjoint.  All of this takes place in wired boundary conditions on a volume $\Lambda$ that is much larger still; the wired boundary conditions represent spins of type 1.  We will be considering the expectation of the random cluster function corresponding to the product of the spin functions.

The important simplifying feature is that, with probability close to one, effective boundary conditions will again isolate the two volumes.  In particular consider the event, $\tilde{\mathcal R}$, that a circuit of occupied bonds separates $V_1$ from $\partial V_2$ {\it and} that this circuit is connected to $\partial \Lambda$; similarly we define $\tilde{\mathcal R}^\prime$.  We claim, on the basis of the arguments in the last two paragraphs in the proof of Proposition A2 that
\begin{equation}
\int \mathbf{P}_{\mathbb J,\Lambda}^{\text{w}}
(\tilde{\mathcal R} \cap \tilde{\mathcal R}^\prime)dF
\to 1
\end{equation}
as first $\Lambda$ and then the $V$'s get large.  Thus, for all intents and purposes, we may consider the random cluster functions under the condition that $\tilde{\mathcal R} \cap \tilde{\mathcal R}^\prime$ occurs.  Noting that these separating rings are part of the boundary cluster (and that we employ the low--temperature modifications described  shortly after Eq.(\ref{product})) we have, for $i = 2, \dots , q$
\begin{equation}
\{A_i\cup B_i \leftrightarrow  \partial \Lambda \}^{c}\cap(\tilde{\mathcal R} \cap \tilde{\mathcal R}^\prime)
=[\{A_i \leftrightarrow \partial V_2 \}^{c}\cap\{B_i \leftrightarrow \partial V^{\prime}_2 \}^{c}]
\cap (\tilde{\mathcal R} \cap \tilde{\mathcal R}^\prime)
\end{equation}
where it has been observed that the under the ring condition, the relevant connections to 
$\partial \Lambda$ occur if and only if the connection to the boundary of the corresponding $V_2$ or $V_2^\prime$ occurs.  Furthermore, among these sets, all disconnection provisos break down to conditions among the $A$'s and $B$'s alone.  I.e.
\begin{equation}
\{A_i\cup B_i \leftrightarrow A_j\cup B_j \}^{c}\cap 
(\tilde{\mathcal R} \cap \tilde{\mathcal R}^\prime) =
\{A_i \leftrightarrow A_j \}^{c}\cap
\{B_i \leftrightarrow B_j \}^{c}
\cap 
(\tilde{\mathcal R} \cap \tilde{\mathcal R}^\prime),
\end{equation}
$2 \leq i, j \leq q$, $i\neq j$.  Thus as far as the indices 2 through $q$ are concerned, under the condition of the ring events, the function is already in product form.  We finally turn attention to the 
term involving the cluster content of $A_1\cup B_1$.  Here it is claimed that with the ``low temperature modifications'' discussed subsequent to Eq.(\ref{product}) {\it and} the condition 
$\tilde{\mathcal R} \cap \tilde{\mathcal R}^\prime$, this term reduces to 
$(q^{-K_W(A_1)}[1 + (q-1){\bf 1}_{A_1 \leftrightarrow \partial V_2}])(
q^{-K_W(B_1)}[1 + (q-1){\bf 1}_{B_1 \leftrightarrow \partial V_2^\prime}])$ i.e~this too is in product form.  The remainder of the proof is similar enough to previous derivations that we need not spell it out in full detail.  The key ingredients are the monotone decreasing property of the wired boundary conditions for all increasing events and the fact (used earlier in a related context) that the connections to increasing boundaries respect this direction of monotonicity.  E.g.
\begin{equation}
{\bf P}_{\mathbb J,V_1}^{\text{w}}(\{A_1 \leftrightarrow \partial V_1 \})
\leq 
{\bf P}_{\mathbb J,\Lambda}^{\text{w}}(\{A_1 \leftrightarrow \partial V_2 \} \mid \tilde{\mathcal R} \cap \tilde{\mathcal R}^\prime )
\leq 
{\bf P}_{\mathbb J,V_2}^{\text{w}}(\{A_1 \leftrightarrow \partial V_2 \})
\end{equation}
where the middle term pertains to conditioning on the ``outermost ring'' in the region 
$V_2 \setminus V_1$.
\qed
\medskip	
	
\medskip

\noindent {\bf Corollary.} {\it  For all temperatures, spin functions of the form $K_{A}  = \prod_{i\in A}\mathbf 1_{\sigma_{i}=1}$ and positive linear combinations thereof are positively correlated in the finite volume and infinite volume} [1]--{\it quenched measures.
Furthermore, the 1--state marginals of these measures conditioned on $\mathbb J$ have positive correlations; i.e.~these marginal measures have ``conditional positive correlations''.  In particular, the above holds in the unique (free) measure associated with the vanishing of the magnetization.}
\begin{proof}
For the type of cylinder function mentioned above, in above mentioned boundary conditions the associated random cluster functions are actually increasing.  Thus, using a double FKG derivation as in Eq.(\ref{double 1}) -- Eq.(\ref{double 2}), we have the positive correlations.
The second statement, while not particularly trivial, was proved in \cite{Chayes} (first Lemma) and generalized in \cite{Hgg3}. 
\end{proof}

\section*{Acknowledgments}
We thank N. Berker for suggesting (to J.L.L.) looking at the disordered Potts models and to  
S. Goldstein for some useful comments. The work of L.C. was supported by the NSF under the grant DMS-0306167  
and the work of J.L.L. and V.M. was supported by NSF grant DMR-044-2066 and AFOSR grant AF-FA 9550-04-4-22910

\end{document}